# Cyclotrons for Particle Therapy

*J.M. Schippers*
Paul Scherrer Institut, Villigen, Switzerland

**Abstract**
In particle therapy with protons a cyclotron is one of the most used particle accelerators. Here it will be explained how a cyclotron works, some beam dynamics aspects, its major subsystems, as well as the advantages and disadvantages of a cyclotron for this application are discussed. The difference between the standard isochronous cyclotron and the synchrocyclotron is explained. New developments are presented and especially those which aim to reduce the size of the accelerator.

## 1      Introduction

In particle therapy, cyclotrons and synchrotrons are the accelerators currently used. The choice depends on treatment method, price, and local conditions, such as available expertise and available space. Synchrotrons are discussed in another chapter in these proceedings. Here, cyclotrons for particle therapy are described, but is important to note that, with both machines, good clinical results have been obtained. The major differences are the footprint of the accelerator, the need for a degrader to set the beam energy of cyclotron beams, and the continuous beam from cyclotrons versus the spill structure of synchrotrons.

Certain characteristics and parameters of the accelerator will depend on the method of the dose-delivery process at the patient. There are two major techniques for applying the dose to the patient. After aiming the beam from the desired direction by setting the gantry (this is a beam-transport system mounted on a rotating mechanical structure), the beam, which has a typical diameter of 1 cm, must be spread in the lateral direction to match the cross section of the tumour as seen from the incoming beam direction. This is done either by the so-called 'scattering technique' or by the so-called 'scanning technique'. In the scattering method, the beam cross section is increased by sending the beam through a scattering system, consisting of one or more foils of material with a high atomic number *Z*, by which the beam diameter is increased to match to the maximum lateral tumour cross section (a 'passive' technique). In the scanning method, a pencil beam of less than 1 cm in cross section is 'actively' scanned in the transverse plane over the tumour cross section. This motion is done in steps and the applied 'spot' dose is varied per step ('spot scanning'), or scanning is performed by continuously shifting the beam along lines in the tumour, during which the beam intensity is varied to deliver the correct dose along the line ('continuous scanning'). Until now, the scattering technique has been used most commonly. However, for several years, the scanning technique has been regarded as the optimal technique (i.e. the technique that applies the best possible dose distribution) currently feasible in practice, and almost all new facilities are designed to employ this technique. Therefore, in this chapter the focus will be on the application of cyclotrons for scanning techniques.

## 2      Basic concept of the cyclotron

Acceleration of charged particles is achieved by means of an electric field. Since electric fields are limited in strength, use is made of the repetitive crossing of electric-field regions in gaps between electrodes. The particles are accelerated when crossing this gap and continue their path through the electrode. Within the electrode, there is no electric field, so the sign of its voltage can be changed without any effect on the particles within. This voltage change must be performed in phase with the particle position; at a phase at which the particle again experiences an accelerating field when leaving the electrode and crosses the gap to the next one. So, only if this voltage change is performed 'in phase', another acceleration step is made. A row of such electrodes is the basic idea of the linear accelerator [1]. Ernest Lawrence investigated the use of a homogeneous magnetic field to bend the particles along a circular trajectory, so that the same gap is crossed repeatedly. The motivation was to reduce the dimensions needed to reach a high energy in a linear accelerator. The centripetal force needed to produce a circular orbit, with radius $r$, of a particle with mass $m$, speed $v$, and charge $q$, is then made by the Lorentz force obtained from the magnetic field $B$:

$$\frac{mv^2}{r} = Bqv. \tag{1}$$

Using this relation, Lawrence realized that the time $T$ a particle needs to make a full circle is

$$T = \frac{2\pi r}{v} = \frac{2\pi m}{Bq}, \tag{2}$$

from which it follows that the revolution frequency $1/T$ does not depend on radius of the particle's circular orbit and, thus, not on the particle energy. All particles have to cross the gap at the same phase, so that they are all, approximately, at the same azimuth (=angular position) in the cyclotron. The frequency at which the electric field in the gap between the electrodes is varying must be in phase with the particle revolution time. In 1929, Ernest Lawrence realized that, since the frequency of the RF signal must match the particle's revolution frequency, it is also independent of the radius and energy of the particle orbit. From Eq. (2), it follows that, for a given particle mass and charge, the RF only depends on the magnetic field. This is the basic principle of the cyclotron operation [2].

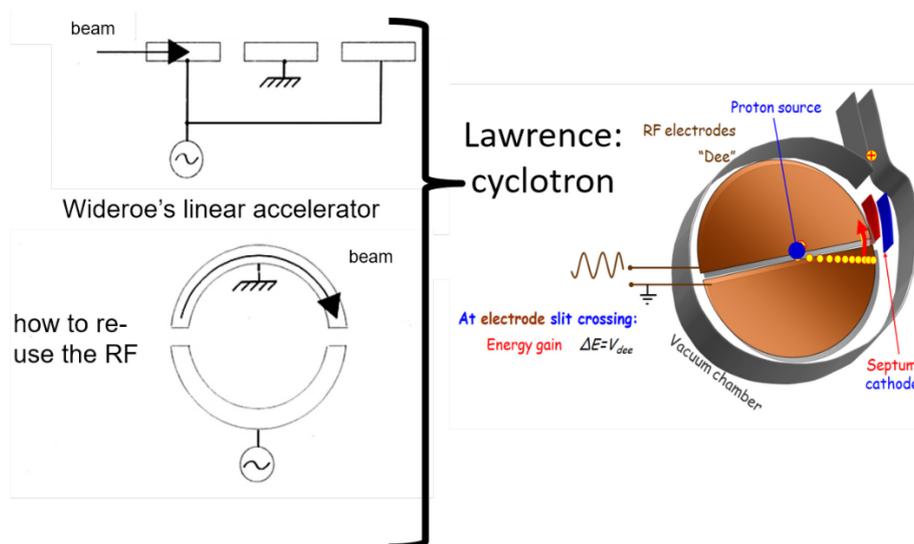

**Fig. 1:** The cyclotron is a combination of the idea of a linear accelerator and a circular version of it, to use the RF repetitively.

As shown in Fig. 1, the major components of a typical cyclotron for therapy are:
– an RF system providing a strong electric field accelerating the protons between electrode plates;
– a strong magnet that confines the particle trajectories into a spiral-shaped orbit, so that they can be accelerated repeatedly by the RF voltage between the electrodes;

- a proton source in the centre of the cyclotron, in which hydrogen gas is ionized and from which the protons are extracted;
- an extraction system that guides the particles that have reached their maximum energy out of the cyclotron into a beam-transport system.

The cyclotron has become a common work horse as an accelerator used in nuclear-physics laboratories and isotope production. The pioneering work of Ernest, and his brother and medical doctor John Lawrence, created interest for the use of energetic beams of heavy charged particles (protons and ions such as He, Ne and C) for therapeutic applications. The 1946 publication of Robert Wilson, showing the advantageous properties of the dose distribution of protons in tissue [3], was a breakthrough in this field. In 1950, John Lawrence was the first to treat patients with cancer by means of a beam of energetic ions [4] from the 60 in. cyclotron in Berkeley (CA, USA), which started its operation in 1939. In 1957, these treatments were successfully duplicated at the cyclotron in Uppsala (Sweden) [5] and, in 1961, at the Harvard cyclotron in Boston (MA, USA) [6]. From then on, the worldwide number of facilities has increased slowly, mostly at facilities located at physics and accelerator laboratories that applied the therapy program in parallel to the physics research. Many cyclotrons, giving proton energies of 60–100 MeV, have been used for treatments of melanoma in the eye. The Harvard Cyclotron was converted to proton-therapy operation in 1949 and has been used solely for proton therapy since 1961. Since then, the group working with this cyclotron has played a major role in the development of proton-therapy techniques. Apart from cyclotrons, synchrotrons also started to play their role in particle therapy. In 1991, the first hospital-based proton-therapy facility came into operation in Loma Linda (CA, USA) [7]. Here, a specially developed synchrotron is used as proton accelerator. The first dedicated, commercially provided, proton-therapy centre using a cyclotron as an accelerator, came into operation at NCC Kashiwa (Japan) in 1998. In 2001, the first patient was treated at the second dedicated, commercially achieved cyclotron, which is based at the proton-therapy facility at the Mass. General Hospital in Boston (MA, USA). This facility was acquired in the first official commercial tender (1992) for a proton-therapy centre. Since then, several commercial companies have been developing cyclotrons for this application. The designs have been based on a simplification of existing cyclotrons by fixing many operational parameters, since therapy requires much less variation of parameter values than applications in physics research. Therefore, only a limited amount of time is needed for tuning a clinically used cyclotron. Furthermore, this simplification increases the reliability, which, together with price, easy operation, and short services, are essential requirements for operation in a clinical environment. In 2007, the first cyclotron using superconducting coils came into operation for a proton-therapy facility at PSI in Switzerland [8]. This Varian machine was the next step in the continuously ongoing process to reduce the size of the accelerator and, thus, the related price of the facility.

Modern cyclotrons, dedicated to proton therapy, accelerate protons to a fixed energy of 230 or 250 MeV. Compared to the classical cyclotrons in accelerator laboratories, the new cyclotrons are rather compact with a magnet height of approximately 1.5 m and a typical diameter between 5 m (200 tons) and 3.5 m (100 tons), when equipped with room temperature coils or with superconducting coils, respectively. Usually, some extra space is needed above and/or below the cyclotron for the support devices of the ion source, RF coupling to the Dees (see Section 3), and equipment to open the machine.

Currently, all operating cyclotrons for particle therapy are accelerating protons. Developments of cyclotrons for acceleration of heavier particles, such as helium or carbon ions, are in progress.

## 3     The RF system

The RF system usually consists of two or four electrodes (often called '*Dee*' due to their shape in the first cyclotrons built) which are connected to an RF generator, driving an oscillating voltage between 30 and 100 kV with a fixed frequency. This RF equals the orbital frequency $1/T$ multiplied by an integer ('harmonic number' $h$, with e.g. $h=2$) and is somewhere in the range of 50–100 MHz. Each Dee consists of a pair of copper plates on top of each other, with a few centimetres in between. The Dees are mounted between the magnet poles, which are at ground potential, as is the whole magnet. When a proton crosses the gap between a Dee and the neighbouring pole, which is at ground potential, it experiences acceleration towards the grounded region when the Dee voltage is positive. When it approaches the Dee at the negative voltage phase, the proton is accelerated into the gap between the two plates. During its trajectory within the electrode or in the ground potential, the electrodes change sign. The magnetic field forces the particle trajectory along a circular orbit, so that it crosses the acceleration gap, the Dees, and the ground, four or eight times during one turn, in the case of an RF system with two Dees or four Dees, respectively.

If the Dee voltage is, for example, 50 kV at the moment of gap crossing in a four-Dee RF system, the proton gains $\Delta E = 0.40$ MeV per turn. Due to the energy gain, the radius of the proton orbit increases, so that it spirals outward. The maximum energy, $E_{max}$, (for proton-therapy cyclotrons, typically, 230 or 250 MeV) is reached at the outer radius of the cyclotron's magnetic field, after approximately $E_{max}/\Delta E = 625$ turns. To limit the number of turns and, thus, the risk of beam losses, it is an advantage to have two, three, or four Dees, so that a high energy gain per turn is achieved at not too high a Dee voltage. Then, the Dees are mounted in the so-called valleys between the pole hills, so that the gap between the poles can be minimized. This pole shape will be explained later.

As mentioned earlier, it is extremely important that the RF is in phase with the azimuthal (angular) position of the particles in the cyclotron. Therefore, such a cyclotron is called an isochronous cyclotron. In a so-called synchrocyclotron, this is achieved by adapting the frequency of the RF signal to a decreasing particle orbit frequency. This will be discussed later.

# 4     Central region of the cyclotron

The protons start their acceleration process in the centre of the cyclotron. In most cyclotrons, an ion source is located here. External sources are of interest, especially when other particles also need to be accelerated. The internal sources operate by exploiting the Penning effect [9, 10]: the ionization of gas by the energetic electrons created in an electrical discharge.

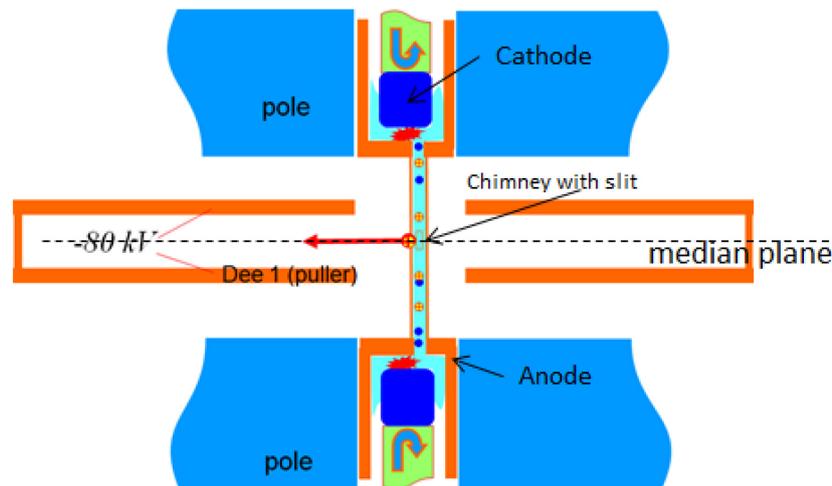

**Fig. 2:** The internal ion source between the magnet poles in the center of the cyclotron

Different configurations of the Penning source are used in cyclotrons, but, in principle, they all operate by the same principles. In the cyclotron at PSI, the ion source consists of two cathodes at a negative voltage of the order of 1 kV, located above and below the median plane of the cyclotron, at each end of a vertical hollow cylinder at ground potential (chimney), see Fig. 2. Hydrogen gas is flushed in between the cathodes and their opposing grounded anodes. Free electrons are created by spontaneous electron emission from the cathodes in the strong electric field between cathode and ground ('cold cathode source'). The electron emission can be stimulated by heating the cathode or by a filament. In the electric field, the electrons are accelerated towards the anodes and they ionize the gas. The ionized gas atoms bombard the cathodes, so that they emit even more electrons. The electrons and ions are confined in gyroscopic orbits along the vertical magnetic-field lines and bounce up and down between the cathodes, thus further ionizing the gas.

The ions ($H^+$, $H_2^+$, $H^-$, etc.) and electrons form a plasma that fills the volume in the chimney between the cathodes. Protons and other ions that have diffused to a little hole in the chimney wall experience the electric field from the nearest Dee edge (the puller). When this Dee is at negative potential, protons that escape from the plasma accelerate towards the Dee. If they arrive at the right phase, they will be accelerated further. Due to the narrow acceptance windows, in time (i.e. RF phase) and in the further acceleration path, only a fraction of the protons leaving the source are actually accelerated.

Important parameters describing the ion source are: the total proton current extracted from the source within not too large an emittance, the stability of the intensity, and the time between services, since these go on cost of operational time of the cyclotron. Modest operational conditions and a careful material choice (heat properties, electron emission, sputtering resistance) are of importance to obtain an acceptably long time interval between source services. Currently, typical time intervals between services are 1–3 weeks.

The maximum beam intensity extracted from the cyclotron (i.e. dose rate at the patient) is determined by the ion-source output. Variation of the source output is done by modifying the gas flow, the cathode voltage, the heating power, or a combination of these. This is a relatively slow process, however, which takes, of the order of, milliseconds. In several scanning processes, there is a need to vary the intensity more rapidly. In addition to adjustment of the source intensity, in several cyclotrons the beam-intensity regulation is performed much faster by intercepting a controlled part of the beam from the source. This is done by a system mounted in the central region of the cyclotron. It consists of an electrostatic deflection plate and intercepting collimators. The source is then operated at a constant intensity, determining the upper limit of the beam intensity extracted from the cyclotron. By deflecting the beam over the slit-shaped collimator aperture, the intensity of the beam passing the aperture can be controlled with high accuracy and within several tens of microseconds. The beam passing the slit will be accelerated. At the slit, the beam energy is still low enough to prevent high power losses and activation.

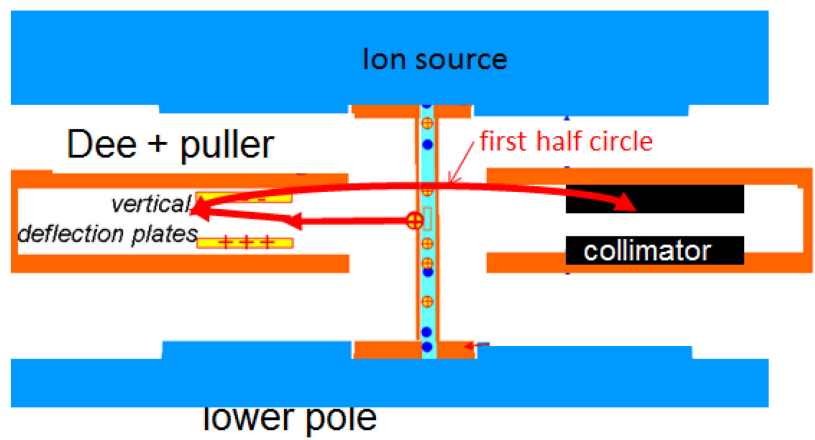

**Fig. 3:** With an adjustable vertical field between vertical deflection plates, the proton beam can be deflected upwards and will be stopped at a collimator jaw. The intensity of the beam, behind the slit between the jaws, is a function of the voltage difference between the vertical deflector plates.

The spiral orbit must be well centred in the magnetic field. A combination of this off-centring and horizontal (radial) betatron oscillations will lead to overlapping turns and to an increase of the horizontal betatron-oscillation amplitude. Due to coupling resonances between horizontal and vertical (axial) betatron oscillations, this may also induce large vertical-oscillation amplitudes. Due to the limited space between the magnet poles and the Dee plates, such large amplitudes could easily cause severe beam losses. Apart from a lower extracted beam intensity, this will also yield neutron production and activation of the cyclotron. This makes the service more complicated.

Centring of the beam is performed by small shifts of the orbit positions, due to a small controlled local variation of the magnetic field in the central region of the cyclotron. This can be done by means of correction coils mounted at the pole surface, or by means of iron pieces that can be shifted, to adjust the pole gap and thus the magnetic field, locally. If one measures the beam intensity as a function of radius in the cyclotron (using a radial probe), an optimal centring is reached by adjusting the magnetic field corrections such that the intensity fluctuations as a function of radius are minimized, as illustrated in Fig. 4.

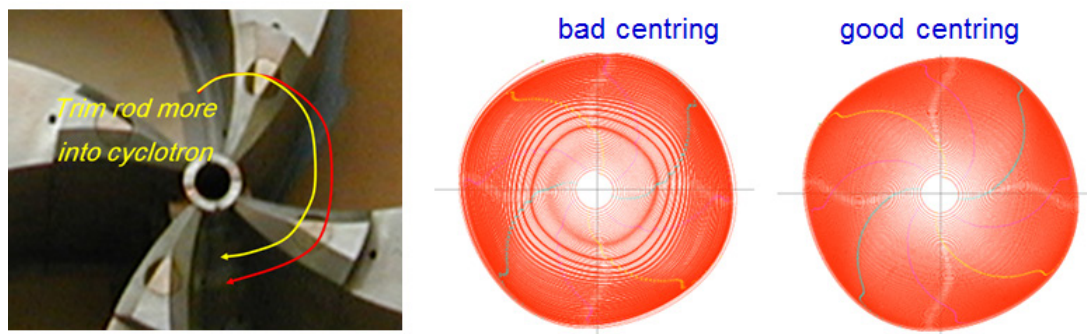

**Fig. 4:** Left: the beam is centred in the first few turns in the cyclotron by adjusting some iron pieces to make some local field fluctuations. A well-centred beam will have a very smoothly varying intensity profile in the radial direction (right). A badly centred beam will show a very irregular radial intensity pattern (middle), since turns with different centres overlap each other at certain radii.

## 5    Vertical focusing during in acceleration in a cyclotron

Vertical focusing is essential during the relatively long distance that that particles have to cover during their acceleration in the cyclotron. Above energies of a few MeV, the focusing due to the electric fields in the acceleration gaps can be neglected. So, focusing is mostly done by the magnetic field. Two vertical focusing processes can be distinguished: *weak and strong focusing*. Weak focusing is used in cyclotrons in which the field decreases as a function of radius. As will be discussed later, often this is the case in cyclotrons that use a very strong magnetic field, necessary to obtain a small machine radius. As shown in Fig. 5, the outward curvature of the field lines causes a component of the Lorentz force in the vertical direction, towards the median plane. It acts on particles that are either above or below the median plane. In most cyclotrons, however, the field is increasing with radius in the cyclotron to compensate the relativistic mass increase of the protons at high energies. This will cause vertical defocusing.

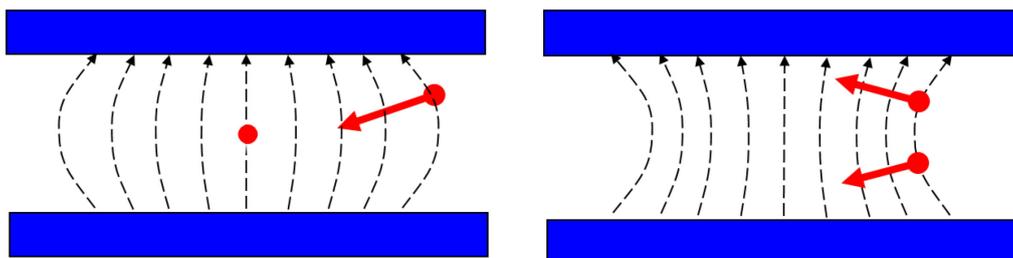

**Fig. 5:** The field lines of the magnetic field and their effect on the vertical focusing. Left: a field which decreases in strength with radius in the cyclotron. Right: a field that increases in strength with radius.

To compensate for vertical defocusing, a stronger vertical focusing is added by an azimuthal variation in the magnet poles by means of 'hills' and 'valleys'. Then, the particles experience a varying field: the field is stronger between the hills and weaker between the valleys. Due to this varying field, the orbits are no longer following an exact circular shape. The orbit has different radii of curvature when crossing a hill or valley, and the boundaries between hills and valleys are crossed non-perpendicular. A non-perpendicular crossing of a change in magnetic field always results in a vertical focusing or defocusing. However, a repeated focusing and defocusing of equal strengths will yield an effective focusing, which is called *strong focusing*. To compensate for the increasing energy with radius and the increasing defocusing by the main field, the angle by which the hill–valley boundary is crossed is increased by making the hill–valley structure spirally shaped, as shown in Fig. 6.

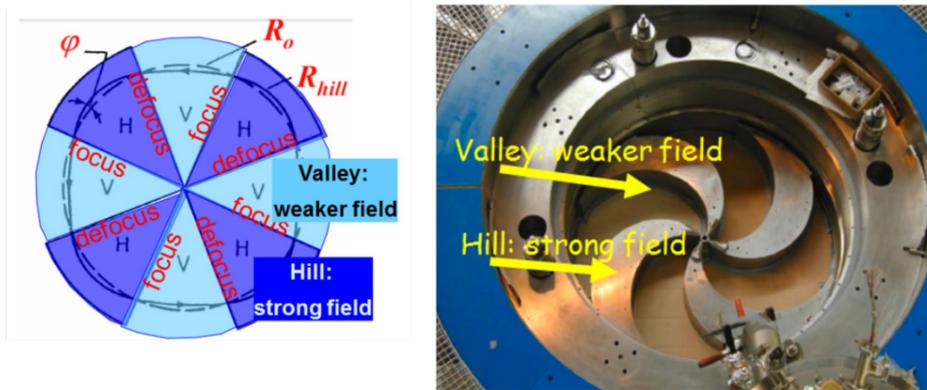

**Fig. 6:** A structure of hills and valleys on the surface of the poles creates an azimuthally fluctuating magnetic-field strength along the almost circular orbit of the particles. This provides the vertical focusing.

## 6   Synchrocyclotron

In order to reduce the diameter of a cyclotron, one must use a stronger magnetic field. This can be achieved with superconducting coils in the magnet. But, as has been mentioned before, in cyclotrons with a very strong magnetic field (4–10 T), the field will decrease as a function of the radius within the cyclotron, due to saturation of the iron at such strong fields and the coil geometry. However, according to Eq. (2), the time to make one turn will increase at lower field strength. Then, the particles will cross the acceleration gaps at too late a phase and will obtain a lower and lower energy increase per turn. At a certain radius, they will be lost. In order to deal with this, the frequency of the RF signal is decreased as a function of time. First, the frequency is matched to the revolution time in the central orbits. Then, the frequency is decreased synchronous to the increasing revolution time at the larger radii. In this way, a group of particles is accelerated from source to extraction, but, in the meantime, particles at other radii cannot be accelerated, since their revolution time does not match with the frequency of the RF signal .This shifting of the frequency of the RF signal is repeated at, approximately, a few hundred Hertz (maximum 1 kHz). As shown in Fig. 7, the beam intensity from such a so-called synchrocyclotron is thus pulsed with this frequency.

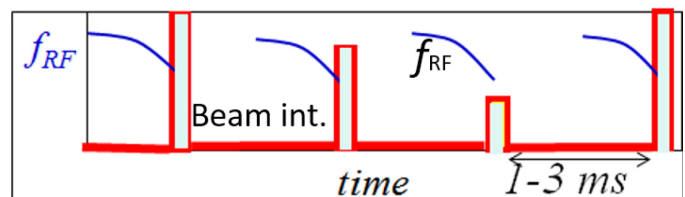

**Fig. 7:** The RF and the beam intensity as a function of time, as extracted from a synchrocyclotron

Currently, several synchrocyclotrons are in development to reduce the size (and price) of a cyclotron for therapy. Since 2010, a synchrocyclotron with a mass of approximately 20 tons, produced by Mevion, has been in operation. It has been mounted on a gantry rotating around the patient [11]. This system is applying the scatter technique, however. In a system recently developed by the company IBA, a small superconducting synchrocyclotron [12] is directly coupled to a spot-scanning gantry in a system as compact as possible. Spot scanning can be performed by applying spots at the frequency at which the frequency of the RF signal is varied. To apply the correct dose, the pulse intensity from the ion source must be set accurately at this frequency (for an ion source this is quite a fast reaction). To obtain sufficient accuracy of the total dose per spot, multiple irradiations of each spot are expected to be necessary.

## 7    Extracting the beam from a cyclotron

When the particles have reached the outer radius of the cyclotron, they have to be extracted from the strong magnetic field. This is achieved by means of a septum: a foil between the last turn in the cyclotron and one of the particles that have to be extracted. At just a few-millimetre-larger radius in the cyclotron, an electrode bar is mounted, parallel to the foil. A strong electric field between the bar at negative potential and the grounded foil pulls the orbit that is to be extracted to a larger radius, where the magnetic field is lower. Using focusing elements, the extracted beam is then guided out of the cyclotron through a channel in between the coils of the cyclotron magnet to the entrance of the beam line.

As can be derived from Eq. (1), the increase per turn of the orbit radius will be smaller at higher energies. Therefore, the orbit separation is rather small at extraction radius and orbits may even overlap. This would cause beam losses at the septum, which should be prevented because of undesired high power loss, activation, and neutron production due to scattered particles. Several methods are used to increase the orbit separation at the extraction septum. The most straightforward methods are to make the cyclotron not too small and to use a high Dee voltage (therefore, a high energy gain per turn). Further, one could exploit extra orbit shifting by a local decrease of the magnetic field, for example, by means of a groove in the pole hill at extraction radius. Another method to increase the radial separation is the so-called resonant extraction, in which a radial betatron resonance is excited. In Fig. 8, it can be seen that the radial spacing between the turns is approximately 1 mm at extraction radius (in this cyclotron at ~0.8 m). But, due to the resonance excited by a small field bump, an orbit has been shifted so that at a certain azimuth, the orbit of next higher energy will be located at a smaller radius. Here, a large turn separation of almost 5 mm occurs at a radius of 81.5 cm, which is of course where the septum should be placed. Using such methods, it is possible to obtain an extraction efficiency (=extracted beam intensity divided by beam intensity in cyclotron centre) of approximately 80%.

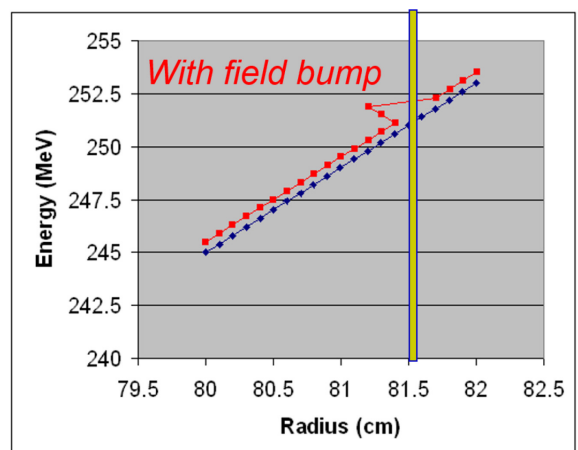

**Fig. 8:** The beam energy as a function of radius. Each point represents a turn. Due to a field bump, a betatron resonance is excited and, at the azimuth where this data is taken, several turns are located at a smaller radial position, due to the orbit shift associated with the excited resonance. Therefore, a large turn separation occurs between the shifted and non-shifted orbits. The location of the septum is indicated.

The energy of the beam at an extraction radius that can be reached in a cyclotron is determined by its magnetic field, its radial profile, and the extraction radius. Although some groups are working on this, currently these parameters are not easy to change sufficiently quickly. Therefore, cyclotrons in therapy are fixed-energy machines. The energy of the beam transferred to the patient is decreased from the energy at extraction to the desired value by means of an adjustable degrader in the beam line, just after extraction. The emittance of the beam extracted from the cyclotron is of the order of a few π mm mrad—usually the vertical emittance is slightly larger than the horizontal emittance. But, since the degrader system increases the emittance so much, the initial emittance is insignificant. Therefore, the asymmetry and exact beam shape of the extracted emittance only play a minor role. Since the emittance of the degraded beam increases with decreasing energy, the transmission of the beam through the beam-transport system to the patient is strongly energy dependent. Therefore, an energy change in the degrader could be coupled to an intensity change of the extracted beam, to get a more or less energy-independent beam intensity at the patient. However, since an error in this method could unexpectedly yield far too high a beam intensity, developments are still in progress in several groups to do this with the ion source and/or vertical deflector plate in the cyclotron.

## 8   Relevant cyclotron characteristics for therapy

When considering a cyclotron for particle therapy, several key parameters need to be considered. A list of these parameters and the issues on which they have an effect, is presented below (Table 1). Here, no numbers are given, since these are extremely dependent on conditions such as how the dose is applied to the patient (i.e. using scattering technique, spot-scan technique, or continuous scanning) and on price.

**Table 1.** Cyclotron parameters that are relevant for therapy applications

| Parameter | Most effect on |
| --- | --- |
| Energy and its stability | Range, sharpness of dose fall off |
| Beam size (emittance) | Transmission to patient, pencil-beam size |
| Beam intensity: structure, stability (kHz), adjustability (range, speed) | Irradiation dose rate, acceptable instantaneous intensity for dosimetry equipment, average intensity—determines the time a treatment takes and affects how spot scanning is done, (im)possibility of continuous scanning |
| Extraction efficiency | Lifetime of components, activation of cyclotron, time needed for service, personal dose |
| Modular control systems + comprehensive user interface | Safety, easy control, fast error tracing, help in decisions in case of 'no beam' |
| Reliability | Number of treatments per year, patient waiting time |
| Activation level (person dose per year) | Personnel safety, time to access machine |
| Maintenance interval, maintenance time, maintenance effort | Number of patients, types of treatments |
| Needed start up time after 'beam off' and after 'cyclotron open' for service | Number of patients, patient waiting time |
| When using ions: time to switch ion species | Number of patients, patient waiting time |
| Choice between synchrocyclotron or isochronous cyclotron | Pulses (<1 kHz) versus continuous beam, scanning possibilities, time a treatment takes |

Both dose application techniques (scattering and scanning) can be applied by means of a cyclotron as the accelerator. However, to improve efficiency of the delivery process when using scanning, it makes sense to provide the cyclotron with several features that can control the beam intensity quickly, accurately, and reliably. For application of the scattering technique, the beam intensity of the cyclotron should not be too low.

## 9  Summary

Currently cyclotrons are used in the majority of proton-therapy facilities. Advantages of cyclotrons are:

- continuous beam (however, pulsed when using a synchrocyclotron);
- 'any' (low or high) beam intensity;
- very quickly and accurately adjustable intensity;
- great reliability (few components);
- relatively small footprint.

But, of course, there are also some disadvantages:

- due to various beam losses the cyclotron gets radioactive, especially when it has low extraction efficiency;
- one and fixed energy, so one needs a degrader in the beam line to set the desired energy;
- activation of components near degrader;
- no carbon ions (yet).

Of course, the accelerator choice strongly depends on many things, such as the type of dose application, available space, local accelerator experience, and, last but not least, the price. It is very interesting to follow the different innovative programs, both in synchrotrons and in cyclotrons.

## Acknowledgement

I would like to thank the numerous colleagues at other particle-therapy institutes and at several companies for the countless discussions at conferences and workshops and the detailed information given at site visits.